\documentclass{aa}

\usepackage[varg]{txfonts}
\usepackage{amssymb}
\usepackage{graphicx}

\usepackage{graphicx}	
\usepackage{mathtools}	
\usepackage{amssymb}	
\usepackage{multicol}   
\usepackage{bm}		
\usepackage{pdflscape}

\usepackage{xcolor}
\usepackage{placeins}

\usepackage{txfonts}
\usepackage{printlen}
\usepackage{soul}

\usepackage{hyperref}

\begin{document} 
   \title{Runaway stars and the Galactic supernova remnant landscape: non-thermal emission and observational evidence}

   \author{R.~Batzofin$^{1}$, 
           K.~Egberts$^{1}$, 
           D.~M.-A.~Meyer$^{2}$,  
           C.~Steppa$^{1}$\\           
          }

    \institute{
    Universit\"at Potsdam, Institut für Physik und Astronomie, 
    Campus Golm, Haus 28, Karl-Liebknecht-Str. 24/25 
    14476 Potsdam-Golm, Germany
    \and 
    Institute of Space Sciences (ICE, CSIC), Campus UAB, Carrer de Can Magrans s/n, 
    08193 Barcelona, Spain
          }
   \date{}

  \abstract
   {   
A significant fraction ($\sim$30\%) of massive stars in our Galaxy are moving supersonically through the interstellar medium, which strongly governs their location at the time they end their lives, e.g. die as a supernova and give birth to a supernova remnant (SNR). 
These dead stellar environments accelerate particles, emitting by non-thermal mechanisms up to the TeV range, and they are considered as a major contributor to the very-high-energy band of the local cosmic-ray spectrum.
   }
   {  
This study investigates the effect of the runaway motion of supernova progenitors on the spatial distribution of SNRs in the Milky Way and how this influences the deduced properties of the population. 
   }
   { 
We construct Galactic populations of SNRs by Monte Carlo simulation, taking into account the bulk motion and the evolution history of their progenitor stars once ejected from their parent clusters. The gamma-ray domain emission of each population is then calculated, to be compared with the High Energy Stereoscopic System (H.E.S.S.) Galactic Plane Survey. 
   }
   { 
We find that including the runaway motion of supernova progenitors strongly modifies the detectability of the simulated emission of their remnants in the very-high-energy band. 

Particularly, our best fit model using a Reid Milky Way model for core-collapse supernova progenitors requires 33\% of massive runaway stars, which is close to the known fraction of runaway high-mass 
stars, to be in accordance with the H.E.S.S. Galactic Plane Survey data.  
   }
   {
Our results show that the runaway nature of supernova progenitors must be taken into account in the study of the Galactic population of SNRs within the H.E.S.S. Galactic Plane Survey and the forthcoming Galactic Plane Survey of the Cherenkov Telescope Array Observatory, as it is a governing factor of the detectability of non-thermal emission of their subsequent SNRs. 
   }


   \titlerunning{Runaway stars and the Galactic supernova remnant landscape}
   \maketitle

\section{Introduction}
\label{intro}

A fraction of massive stars are moving supersonically through their ambient medium~\citep{blaauw_bain_15_1961,
last_paper_on_runaway_stars}. The fraction spans from a few to 50\% \citep{Kobulnicky_2022} but recent studies converge to $\sim$30\%. This movement comes mainly from the ejection of stars from their parent clusters due to gravitational swing, however, there are other mechanisms for this, such as ejection from multiple systems~\citep{Hoogerwerf_2000ApJ...544L.133H, gvaramadze_mnras_410_2011,bromley_apj_706_2009}. Such motion does not prevent stellar evolution, and while running over pc-scale distances, 
the stars keep on undergoing profound changes in their inner core reflecting in the bow shock circumstellar nebulae~\citep{weaver_apj_218_1977,gull_apj_230_1979,wilkin_459_apj_1996} and leading to a final supernova explosion~\citep{woosley_araa_24_1986}. Consequently, a significant proportion of core-collapse (CC) supernova remnants (SNRs), originating from such runaway stars, are displaced with respect to their progenitor's birthplace and might explain the existence of high Galactic latitude remnants~\citep{katsuda_apj_863_2018}.

There is evidence that SNRs accelerate particles, however, it is unknown whether they are the sources of all Galactic cosmic rays, this is a long standing open problem in astroparticle physics. The standard paradigm is that they accelerate particles via diffusive shock acceleration. The accelerated protons and electrons interact with the interstellar medium to produce gamma rays in the very-high-energy domain, predominantly via (i) pion decay from proton-proton interactions (hadronic) and (ii) inverse Compton scattering of electrons on soft photons (leptonic). Galactic population studies have been performed, comparing simulated populations of SNRs in the TeV range by their gamma-ray emission with TeV measurements. These optimisations do not yield unique results due to degeneracies in the  parameter space. So far, these works have assumed that the progenitors of supernovae were in-situ objects \citep{1st_paper}.

SNRs have been observed in many wavelengths from radio to gamma rays, and this information is gathered in the Green catalogue\footnote{\tiny Green D. A., 2024, ‘A Catalogue of Galactic SNRs (2024 October version)’, Cavendish Laboratory, Cambridge, United Kingdom (available at \url{https://www.mrao.cam.ac.uk/surveys/supernova remnants/)}}
\citep{Green_2019, Green_2022, Green_2024}. The catalogue documents all known Galactic SNRs and was last updated in October 2024. At the time of writing, there are 310 SNRs in the catalogue, most of which have been detected and identified by radio instruments. 
The catalogue provides positional information such as the objects' Galactic latitudes. In addition, \citet{Ranasinghe_2022} published distances for 215 SNRs.

The questions are therefore: How does the runaway nature of supernova progenitors influence the SNR population of the Milky Way? How does this modify the TeV detectability and the reconstructed non-thermal emission properties of the SNR population?

This paper includes the runaway nature of stars as supernova progenitors to the distribution of SNRs. We investigate its effect on the spatial distribution, confront the simulations with the H.E.S.S. data from the Galactic Plane Survey (HGPS) \citep{hgps_2018} and the Green catalogue for SNRs \citep{Green_2024} and derive the emission properties in the TeV range of a Galactic population model. In section~\ref{method} we describe our runaway progenitor model, in section~\ref{results} we present and discuss our results and we draw our conclusions in section~\ref{discussion}.

\section{Method}
\label{method}
We constructed CC and thermonuclear pre-supernova population models following the \citet{Reid_2019} distribution based on evolving massive stars forming in the Galactic plane and the \citet{Steiman-Cameron_2010} distribution following the [CII] lines of the interstellar medium, respectively. These progenitors are potentially dynamically ejected from the plane. The distribution of the subset of runaway stars as a function of their bulk velocity is given by $dN(v_\star) \propto e^{-v_\star/v_{\rm max}}$, with $v_{\rm max} = 150 \rm km \:s^{-1}$ \citep{bromley_apj_706_2009}. Each star is assigned two random angles (azimuthal and polar) for the direction of the motion, assumed to be a straight segment of size, $v_\star t_{\rm SN}$, where $t_{\rm SN}$ is the lifetime of the star. Their initial mass, given by the initial mass function~\citep{kroupa_mnras_322_2001}, governs their evolution~\citep{ekstroem_aa_537_2012}. We refer the reader interested in the details on the distances travelled by the stellar population to App.~\ref{appendix_runaway}. The emission of each SNR is estimated based on particle acceleration \citep{cristofari_2013, Ptuskin_2003, Ptuskin_2005} and reacceleration \citep{Cristofari_2019}. The gamma-ray emission is calculated using the {\sc naima} code \citep{naima} using the pion decay model \citep{naima_PD} and the inverse Compton model \citep{naima_IC}. More details of the population model can be seen in App.~\ref{appendix_pop_model}. The gamma-ray luminosity, size and position of each SNR is used to determine its detectability in the HGPS, following the method of \citet{Steppa_2020}.  When comparing the number of detectable simulated SNRs with the HGPS we use lower, stringent, and strict limits. The lower limit (8) is the firmly detected SNRs, the stringent limit (28) is the lower limit plus the composite objects (8) plus the unidentified sources associated (by position) with a SNR (12), and the strict limit (63) is the stringent limit plus the rest of the unidentified sources (35). The three main parameters explored in our model are the spectral index of accelerated particles ($\alpha$), the efficiency of particle acceleration ($\eta$) and the electron-to-proton ratio ($K_{\rm ep}$). Using a static source distribution based on \citet{Steiman-Cameron_2010}, we found that the parameter combination with the most populations within our stringent limits was $\alpha = 4.2$, $K_{\rm ep} = 10^{-5}$, $\eta = 0.09$.

\section{Results}
\label{results}

\subsection{
A Galactic population of runaway massive stars induces less supernova 
remnants detectable by the HGPS
}
\label{finding1}

\begin{figure*}[h]
    \centering
    \includegraphics[]{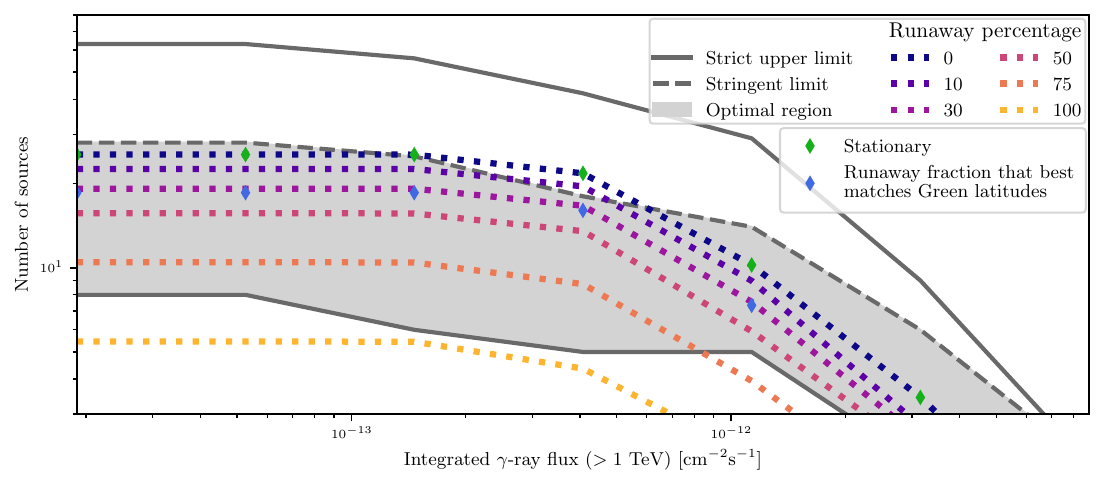}
    \caption{
    Distribution of the number of simulated SNR sources as a function of the 
    integrated gamma-ray flux above 1 TeV observable in the HGPS. The grey region shows the ideal region for the simulations: between the identified SNRs and the sources associated with SNRs. The coloured dotted lines are the simulated populations assuming different percentage of runaway massive stars progenitors of the CC SNRs. The diamonds indicate the stationary (matching 0\% runaways) and the fraction of runaways that best fit the latitudes in the Green catalogue. 
    }
     
    \label{fig:logNlogS_run_with_best_fits}
\end{figure*}

Including a subset of runaway stars in the Galactic population model of~\citet{1st_paper}, has the following effect: An increase in the percentage of runaway CC SNR progenitors naturally decreases the number of simulated SNRs that would be detectable in the HGPS. This is in part because an equivalent fraction of the gamma-ray bright SNRs move out of the region covered by H.E.S.S. observations ($|b| \leq 3^\circ$), and in part because the interstellar medium is less dense further from the plane, which causes the SNRs to be less luminous and larger in size, making them harder to detect. We investigated the impact of changing the runaway percentage using the initial parameters that we identified previously as the best match to the HGPS data. We set the thermonuclear runaway percentage to 0\% for this part. In Fig.~\ref{fig:logNlogS_run_with_best_fits}, the $\log N - \log S$ distribution of detectable SNRs is shown (including both thermonuclear and CC SNRs). The change in detectable SNRs as the CC progenitor runaway percentage changes is shown by the coloured dotted lines. 
Changing from no runaways to all the CC SNRs coming from runaway progenitors, the number of HGPS-detectable sources for the given parameter set drops from 25.37 to 5.45, on average for the 100 populations. In comparison with the limits obtained by the HGPS data (grey shaded region), no runaway stars are at or above the stringent upper limit, while all progenitors being runaway stars is just at or below our lower limit, respectively. This is our first result, i.e. runaway massive stars affect the detectability of HGPS SNRs.

\subsection{
The current observed population of CC SNRs of the Green catalogue 
is in agreement with the known proportion of runaway massive stars
}

We constrained the proportion of runaway massive stars in the Milky Way, which is known from the literature to be $\sim$30\% of the stellar population in the high-mass regime. We fitted the percentage of runaways required to best match the latitude distribution of the SNRs in the Green catalogue, which is independent of the HGPS campaign. As the latitude distribution of an incomplete catalogue crucially depends on the distance of the detected sources (for simple geometrical reasons), we select only those SNRs from the Green catalogue whose distance is known \citep{Ranasinghe_2022} and choose from our simulated populations sources that match the distance distribution of the Green catalogue sources, thus mitigating the bias. A two-sample Kolmogorov-Smirnov test \citep{KS_test} is used to determine the similarity of the latitudes between our simulated sources and the sources in the Green catalogue. For details of the fitting, see appendix~\ref{appendix_Green}. The best fit has a CC runaway percentage of 33\% and a thermonuclear runaway percentage of 0.02\%.

We investigated the latitude distributions of simulated SNRs with sources from the HGPS by varying the fraction of CC runaway progenitors from $0\%$ to $100\%$, again the thermonuclear runaway percentage is set to 0\%. We categorise the HGPS sources into three groups based on their identification and use the two-sample Kolmogorov-Smirnov test to assess similarity. The categories follow our lower, stringent, and upper limits; their latitude distributions are shown in Fig.~\ref{fig:HGPS_latitudes}. 
We find that as the runaway percentage increases, so does the similarity between the distributions of the HGPS SNRs and the simulated SNRs. A challenge in the comparison of simulations and data is the asymmetry of the HGPS-detected sources with a shift to negative latitudes, which is not reproduced by simulations. A less pronounced effect is also visible in the latitude distribution of the Green-catalogue SNRs.

This result provides insight into the progenitor distributions and potential selection effects within the Green catalogue by constraining the percentage of runaway massive stars to 33\% in the Galactic population, in accordance with the literature.

\subsection{
Our proportion of runaway massive stars constrained from the Green catalogue 
latitude distribution produces a supernova population distribution in accordance 
with the observed HGPS $\log N - \log S$ distribution
}

We calculated the $\log N - \log S$ distribution as would be observable within the HGPS for a Galactic population of SNRs, accounting for the fact that a subset of the CC SNRs, previously constrained to 33\%, originate from massive runaway stars. We explore the parameter space for this new source distribution using the work flow of~\citet{1st_paper} and find that the parameter sets that have the highest number of populations in agreement with our stringent limits, while also including at least four sources with $E_{\rm max} > 10$ TeV, includes amongst others the same as in~\citet{1st_paper} ($\alpha = 4.2$, $K_{\rm ep} = 10^{-5}$, $\eta = 0.09$, $\alpha = 4.15$, $K_{\rm ep} = 10^{-5}$, $\eta = 0.05$, and $\alpha = 4.2$, $K_{\rm ep} = 10^{-4.5}$, $\eta = 0.07$). Notably, 100\% of these populations are in agreement with the HGPS. As in our previous work, multiple parameter sets produce a large fraction of populations that align with the HGPS constraints. The log $N$–log $S$ plot for the best-fit parameters, both for the run percentage that best matches the latitudes from the Green catalogue and for the case with no runaway SNRs, is shown in Fig.~\ref{fig:logNlogS_run_with_best_fits}.

\begin{figure}[h]
    \centering
    \includegraphics[]{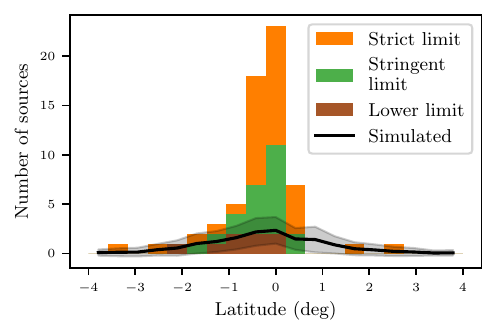}
    \caption{
    Latitude distribution of the HGPS sources, 
    split up into their source types and the simulated sources, with a runaway fraction of 100\% for CC progenitors and 0\% for thermonuclear progenitors, and the following parameter set: $\alpha=4.15, \: K_{\rm ep} = 10^{-4.5},\: \eta = 0.07$.
    }
    \label{fig:HGPS_latitudes}
\end{figure}

Using the extreme case of all CC SNRs originating from runaway progenitors as suggested by the HGPS latitude distribution, we find two sets of parameters within the formalism of ~\citet{1st_paper}, with 100\% of the populations within the stringent limit and having at least 4 SNRs with $E_{\rm max} > 10 \rm$ TeV. In accordance with our first finding of section~\ref{finding1} and Fig.~\ref{fig:logNlogS_run_with_best_fits}, a 100\% fraction of runaways requires a larger number of detectable SNRs, therefore modifying the optimal parameter set with respect to the stationary case to favour brighter sources. The optimal parameter sets are: $\alpha=4.15, \: K_{\rm ep} = 10^{-4.5},\: \eta = 0.07$ and $\alpha=4.05, \: K_{\rm ep} = 10^{-5},\: \eta = 0.03$.

The latitude distribution of an example parameter set of $\alpha = 4.15,\: K_{\rm ep} = 10^{-4.5},\: \eta = 0.07$ is shown in Fig.~\ref{fig:HGPS_latitudes}.
Comparing the latitudes of the detectable simulated SNRs from these parameter sets to the latitudes of the sources in the HGPS we obtain the following p-value ranges: 
$0.28^{+0.24}_{-0.14}$ (SNRs), $0.22^{+0.25}_{-0.16}$ (SNRs and composites), $0.16^{+0.22}_{-0.12}$ (all associated with SNRs), $0.11^{+0.14}_{-0.09}$ (including all unidentified sources). 
It is possible to match both the latitude distribution and the flux distribution of the HGPS sources when assuming the extreme case of all CC SNRs coming from runaway progenitors.

We conclude that the fraction of runaway massive stars, as inferred from the latitude distribution of the Green catalogue, leads to a supernova population distribution that aligns with the observed HGPS log $N$–log $S$ distribution. A larger fraction of runaways as indicated by the HGPS latitude distribution requires a brighter SNR population; this can be achieved by increasing the electron fraction, having a harder spectral index or increasing the acceleration efficiency. 

\section{Discussion and conclusion}
\label{discussion}
We investigated the effects of the bulk motion of massive stars onto the Galactic non-thermal emission as observed by the HGPS. We highlight that a subset of moving massive stars in the Milky Way implies that less SNRs are able to be detected by the HGPS survey. Furthermore, we constrained this proportion of runaway high-mass stars by Monte-Carlo simulation that we tested against the current observed population of SNRs in the Green catalogue and found that the runaway percentage is about 33\% for CC progenitors and about 0.02\% for thermonuclear progenitors, which is well in agreement with the known proportion of runaway massive stars in the literature. Including this percentage of fast-moving supernova progenitors in the population synthesis of ~\citet{1st_paper} modifies the simulated log $N$–log $S$ distribution of HGPS-detectable sources. We found optimised parameter sets for that simulated log $N$–log $S$ distribution, which both include this runaway stars percentage and produce a supernova population distribution whose emission is in accordance with the observed HGPS log $N$–log $S$ distribution. Comparing our simulations with the latitude distribution of HGPS sources yields significantly different estimates, namely suggesting all CC SNRs to originate from runaway progenitors and favouring also different parameter sets describing the population properties. Although the divergence of these results persists, it has to be noted that the pronounced asymmetry in the H.E.S.S. latitude distribution challenges the detailed description with a symmetrical model. Despite these challenges, we are able to exclude large portions of our parameter space, such as $\alpha \geq 4.35$ and $K_{\rm ep} > 10^{-3}$.
Our study concludes that including the bulk motion of massive stars, given its governing effect onto the distribution of SNR within the latitudes of the Milky Way, is an element that is necessary to take into account for characterising the Galactic supernova-remnant population.

Beyond these conclusions, one may ask whether the factors governing the ejection of high-mass runaway stars might, 
in turn, influence our results. These factors include both the spatial distribution of massive star-forming 
regions within the Milky Way and the mechanisms responsible for stellar ejection from clusters containing 
massive stars~\citep{Hoogerwerf_2000ApJ...544L.133H, Schoettler_2019MNRAS.487.4615S}. The former arises from the star formation history of the Galaxy, while the latter involves detailed N-body 
dynamics that lead to the ejection of massive stars from their natal environments. In other words, the Galactic 
star formation rate, the initial mass function \citep{kroupa_mnras_322_2001}, and the intrinsic physics of 
stellar clusters \citep{kraus2020} drive local variations in the distribution of young massive stars — factors that are not incorporated into our current model. The effect of binary interactions on the stellar evolution \citep{Marchant_2024} can be assumed to play a minor role for our purposes and is neglected.
Additionally, the motion of stars can be expected to further influence the compression ratio of the 
SNR’s expanding shock front, particularly when interacting with the progenitor's circumstellar medium, which will affect particle acceleration processes and the resulting non-thermal emission associated 
with these remnants \citep{meyer_mnras_521_2023,meyer_aa_687_2024}.

More observations, particularly of higher-latitude SNRs, would help to narrow down the expected number of SNRs forming from runaway progenitors. Future instruments, such as the Cherenkov Telescope Array Observatory (CTAO) \citep{GPSCTA} and the Large High Altitude Air Shower Observatory (LHAASO) \citep{LHAASO_SNRs}, should greatly increase the number of detected SNRs, helping to refine our population model further, understand the physics of massive star formation and the functioning of stellar ejection out of massive stellar clusters, to better constrain the runaway nature of CC progenitors and their effects on the properties of the Galactic supernova-remnant population.

\begin{acknowledgements}

The authors would like to give the referee a big thank-you, their suggestions and comments significantly improved the work. The authors are grateful for the computing time on the high-performance computer "Lise" at the NHR Center NHR@ZIB, jointly supported by the Federal Ministry of Education and Research and the state governments participating in the NHR (www.nhr-verein.de/unsere-partner). This work is funded by the Deutsche Forschungsgemeinschaft (DFG, German Research Foundation) with the grant 500120112 and supported by the grant PID2021-124581OB-I00 funded by MCIN/AEI/10.13039/501100011033 and 2021SGR00426 of the Generalitat de Catalunya. This work is also supported by the Spanish program Unidad de Excelencia Mar\'ia 
de Maeztu CEX2020-001058-M. This work is also supported by MCIN with funding from European Union NextGeneration EU (PRTR-C17.I1).

\end{acknowledgements}

\bibliographystyle{aa}
\bibliography{runaway_stars}

\begin{appendix}
\section{Distance travelled for runaway progenitors}
\label{appendix_runaway}

\begin{figure}[h]
    \centering
  
    \includegraphics[]{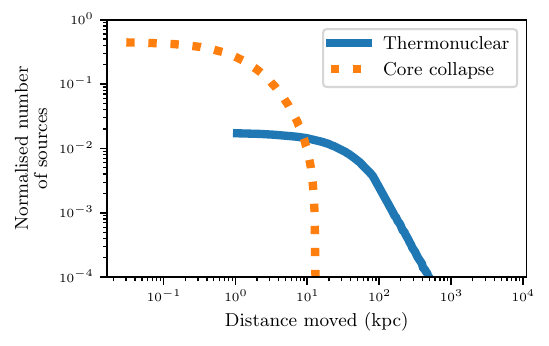}
    \caption{
    Normalised distribution of distances travelled by the progenitor star of supernova 
    remnants in the Milky Way, for both thermonuclear (blue, solid line) and CC 
    (orange, dotted line) progenitors. 
   }
    \label{fig:distance_dstribution}
\end{figure}

The distance that a runaway progenitor travels is calculated using the velocity and lifetime of the progenitor.bulk velocity is given by $dN(v_\star) \propto e^{-v_\star/v_{\rm max}}$, with $v_{\rm max} = 150 \rm km \:s^{-1}$ \citep{bromley_apj_706_2009}. The parameter needed for this calculation is the lifetime of the star $t_{\rm SN}$: For CC supernovae, we start by considering the high-mass regime of the initial mass function, i.e. the distribution mass function of massive stars when reaching the onset of their main-sequence in a given star-formation region~\citep{kroupa_mnras_322_2001}, to get a mass distribution for the CC SNR progenitor, see also the method in~\citet{1st_paper}. The relation between the zero-age main-sequence masses of high-mass stars and their lifetimes is estimated on the basis of tabulated evolution models interpolated from the {\sc geneva} library\footnote{https://www.unige.ch/sciences/astro/evolution/en/database/syclist/} \citep{2008Ap&SS.316...43E,ekstroem_aa_537_2012}. This code calculates one-dimensional stellar structures including internal convection, diffusion~\citep{zahn_aa_256_1992} and mass-loss physics~\citep{Georgy_2010PhDT}. The final age, $t_\mathrm{SN}$, is estimated as the moment the inner core of massive stars triggers Si burning. Thermonuclear SNRs are much less likely to be formed from runaway stars ($< 2\%$), however, they can travel for a much longer time through the interstellar medium, and consequently, finish their life further from the Galactic plane than higher-mass stars. In order to determine $t_{\rm SN}$ for the thermonuclear supernova progenitors, we use the delay time before the explosion. The distribution of delay times derived from the basis of simulations is the power law  $f(t) \propto 100 t^{-1}$, see \citet{Ruiter_2009}. 
The distance travelled by a star $v_{\star}t_{\rm SN}$ is then obtained, see Fig.~\ref{fig:distance_dstribution}.

\section{SNR population model}
\label{appendix_pop_model}
The SNR population model has three ingredients: the physics of the SNR, the spatial distribution, and the matter distribution. The matter distribution follows \citet{Shibata_2010}, which has an empirical model for the gas density of hydrogen ($H_1$ and $H_2$) as a function of galactocentric distance and height above or below the Galactic plane.

The source distribution starts with the the CC SNR distribution following \citet{Reid_2019}, which is based on massive stars with maser parallaxes and the thermonuclear SNR distribution following \citet{Steiman-Cameron_2010}, which is based on measurements of the interstellar medium ([CII] line).
  This stationary distribution is then altered by some of the SNRs being runaway and not remaining in the original position.

We assume that 32\% of the SNRs are thermonuclear and 68\% are CC. The thermonuclear SNRs have an explosion energy of $10^{51}$ erg and the mass of the ejecta is $1.4 M_\odot$. For CC SNRs, we use the high-mass regime of the initial mass function to get the distribution of initial masses from which we estimate the ejecta mass and explosion energy of the SNR. The evolution of the SNRs is the same as in \citet{1st_paper}. The dynamical evolution of the shock, the shock radius and velocity as a function of time, is computed in the same way as in \citet{Ptuskin_2003, Ptuskin_2005, Chevalier_1982, Truelove_1999}, taking into account the density of interstellar matter at the location of the SNR. In addition the magnetic field amplification due to non-resonant streaming is taken into account. We assume that some fraction $\xi_{\rm CR}$ of the ram pressure is converted into cosmic ray protons at the shock; and that a fraction $K_{\rm ep}$ of $\xi_{\rm CR}$ is converted into accelerated electrons. The gamma-ray emission is calculated from the spectrum of protons and electrons, assuming an electron-proton ratio and a spectral index.

The only changes in the population model from \citet{1st_paper} is in the distribution of SNRs. We now separate the distributions of the CC and thermonuclear SNRs and include runaway SNRs for both types of SNRs.

\section{Fitting the runaway percentage using the SNRs in the Green catalogue}
\label{appendix_Green}
\begin{figure}
    \centering
    \includegraphics[]{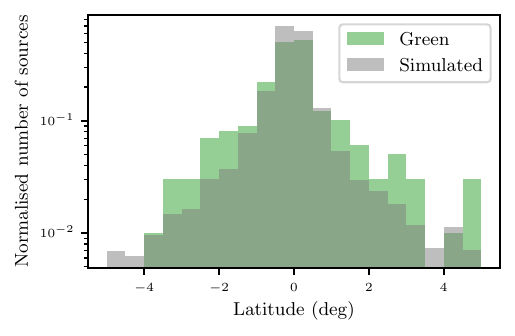}
    \caption{
    Histogram of the latitudes of Green SNRs and the simulated SNRs chosen to match the distance distribution of the Green catalogue, with a runaway fraction of 33\% for CC progenitors and 0.02\% for thermonuclear progenitors.
    }
    \label{fig:Green_hist}
\end{figure}

We restricted the maximum thermonuclear SNR runaway percentage to 2\%, while allowing the CC SNR run percentage to vary between 0\% and 100\%. The simulated SNRs in each population are selected such that the distribution of the distances in the simulated population is matching to that of the SNRs in the Green catalogue. Using the two-sample Kolmogorov-Smirnov test we found that the simulated populations whose SNR latitudes most closely match those of the Green catalogue sources have a CC runaway percentage of 33\% and a thermonuclear runaway percentage of 0.02\%, the p-value for this selection is 0.37. Thus, we cannot reject the null hypothesis that the latitude distributions of the simulated and observed populations are statistically matching, suggesting that the selection process effectively minimises biases and that the simulated populations reasonably represent the observed distribution of SNRs. The initial positions of the SNRs naturally have an impact of the runaway percentage that best matches the Green data. We found that with different distributions as much as 50\% CC progenitor runaways are required to best match the Green data. In Fig.~\ref{fig:Green_hist} the simulated latitudes are shown with the Green latitudes, taking into account the distance distribution, showing the similarities of the two distributions. 

\end{appendix}

\end{document}